\documentclass[12pt,3p,authoryear]{elsarticle}
\usepackage[T1]{fontenc}
\usepackage{amsmath,amssymb,amsfonts}
\usepackage{natbib}
\usepackage{graphicx}
\usepackage[hyphens]{url}
\usepackage[table]{xcolor}
\usepackage{epstopdf}
\usepackage{booktabs}
\usepackage{rotating}
\usepackage{times}
\usepackage{bm}
\usepackage{xspace}
\usepackage{color}
\usepackage{subdepth}
\usepackage{float}
\usepackage{setspace}
\usepackage{needspace}   % almost always useful
\usepackage[labelfont=bf,position=top]{caption}
\usepackage{geometry}
\usepackage[%
pdftex,%
pdfauthor={Me}%
pdfstartview={XYZ null null 1},%
pdffitwindow=true,%
pdfcreator={pdflatex}
letterpaper,%
colorlinks,%
breaklinks=true,%
backref=true,%
plainpages=false,%
bookmarks=false,%
pdfpagelabels,% this may add another page
pdfpagemode=None%
]{hyperref}

\newcommand\CB[1]{\colorbox{gray!40}{#1}}

\newcommand\CBX[1]{\colorbox{gray!20}{#1}}

\usepackage[font={small,singlespacing},labelfont={bf,small},skip=0.2cm]{caption}
\floatstyle{plaintop}
\restylefloat{table}
\restylefloat{figure}

\RequirePackage{ae,fancyvrb}
\newcommand{\pkg}[1]{\textbf{#1}}
\newcommand{\R}{\textsf{R}\xspace}

\journal{arXiv}

\usepackage{adjustbox}
\usepackage{longtable}
\usepackage{tabularx,xltabular}

\newcommand{\ie}{\emph{i.e.}\xspace}
\newcommand{\eg}{\emph{e.g.}\xspace}

\usepackage{amsmath}% http://ctan.org/pkg/amsmath
\usepackage{kbordermatrix}% http://www.hss.caltech.edu/~kcb/TeX/kbordermatrix.sty

\renewcommand{\kbldelim}{(}% Left delimiter
\renewcommand{\kbrdelim}{)}% Right delimiter

\DeclareMathOperator*{\argmin}{arg\,min}

\newcommand{\insertfloat}[1]{%
\begin{center}
[Insert~#1 about here.]%
\end{center}%
}

\begin{document}
\begin{frontmatter}
\title{Optimal Text-Based Time-Series Indices}
\author[hec]{David Ardia}
\ead{david.ardia@hec.ca}
\author[udes]{Keven Bluteau\corref{cor1}}
\ead{keven.bluteau@usherbrooke.ca}
\address[hec]{GERAD \& Department of Decision Sciences, HEC Montréal, Montréal, Canada}
\address[udes]{Department of Finance, Université de Sherbrooke, Canada}
\cortext[cor1]{Corresponding author. }

\begin{abstract}
We propose an approach to construct text-based time-series indices in an optimal way---typically, indices that maximize the contemporaneous relation or the predictive performance with respect to a target variable, such as inflation. We illustrate our methodology with a corpus of news articles from the Wall Street Journal by optimizing text-based indices focusing on tracking the VIX index and inflation expectations. Our results highlight the superior performance of our approach compared to existing indices.
\end{abstract}
\end{frontmatter}

\doublespacing
%\onehalfspacing
%\singlespacing

\newpage
\section{Introduction} 

In economic and financial research, there is a growing trend of integrating textual data such as news articles into econometrics analysis \citep[see][for a review]{gentzkow2019text}. This integration is typically done by (i) selecting, (ii) transforming, and (iii) aggregating textual content into a time-series representation \citep[see][for a general overview of these steps]{ArdiaEtAl2019,algaba2020econometrics}. While many studies have focused on steps (ii) and (iii)---transforming and aggregating textual data into a quantitative measure such as sentiment \citep[see \eg,][]{loughran2014measuring,jegadeesh2013word,manela2017news}---the essential selection step (i), which usually relies on subjective ad-hoc rules, has not received much attention yet. 

We aim to fill this gap in this article by proposing an approach to construct text-based time-series indices optimally. Specifically, our algorithm determines which set of texts, among a large corpus, leads to a text-based index that is optimal for a specific objective---typically, an index that maximizes the contemporaneous relation or the predictive performance with respect to a target variable, such as inflation. Our methodology relies on binary \emph{selection matrices} that, applied to the vocabulary of tokens, select the relevant texts in the corpus. Various widely-known text-based indices, such as the Economic Policy Uncertainty (EPU) index by \citet{BakerEtAl2016}, can be formulated in terms of selection matrices. 

Optimizing selection matrices is challenging due to the inherent non-linearity that arises when aggregating selected texts into text-based indices. To overcome this difficulty, we design a genetic algorithm with domain-specific knowledge to explore the solution space and obtain the \emph{optimal} selection matrix. The algorithm starts with an initial population of selection matrices. These matrices are evaluated using a fitness function that measures the resulting textual-index performance in achieving the objective of selecting the texts. At each iteration, the population of selection matrices undergoes tailor-made crossover and mutation operations, as traditional operators are not well suited to this optimization problem. We also implement additional steps to address potential overfitting issues and leverage the information in word embeddings of promising solutions to help explore good solutions more efficiently. Finally, we introduce a pruning step to avoid sub-optimal solutions. 

To showcase the relevance of our methodology, we conduct two empirical applications using a collection of 763,542 Wall Street Journal news articles spanning from January 2000 to August 2021.

First, we validate our methodology with the EPU index. Specifically, we use our algorithm to see if we can recover the set of keywords proposed by \citet{BakerEtAl2016} when building their EPU index. Given the number of tokens in our corpus vocabulary, recovering the set of tokens for the three dictionary dimensions of \citet{BakerEtAl2016} is not trivial. We show that our approach (i) can recover the set of keywords proposed by the authors---their selection matrix---and (ii) does so in a reasonable computational time. 

Second, we evaluate the performance of our methodology against established benchmarks, focusing on the monthly VIX index and inflation expectations derived from the Michigan Surveys of Consumers. In this comparative analysis, the text-based indices generated by our methodology consistently outperform the text-based benchmarks on the test window, a time period not considered during the optimization of the selection matrices.

The rest of this paper is organized as follows. Section \ref{sec:selection} introduces the notation, presents the concept of selection matrices, and outlines the optimization problem. Section \ref{sec:optimization} presents our optimization strategy, including the genetic algorithm as well as the proposed crossover and mutation operators. Section \ref{sec:empirical}  presents our empirical applications, and Section \ref{sec:conclusion}  concludes. 

\section{Token-Based Text  Selection}
\label{sec:selection} 

We first introduce the concept of tokens and vocabulary to analyze a corpus of texts. A token, denoted by $v$, represents a sequence of characters, including acronyms, words, sequences of words, or even regular expressions. The vocabulary of size $V$ is defined as a collection of such tokens.

The text corpus is represented as a matrix $\bm{C}_t$ of size $N_t \times V$, where $N_t$ is the number of texts available at a given time $t$. Each element $c_{n,v,t}$ among a collection of matrices $\bm{C}_t$, where $t = 1,\dots,T$, indicates if the token $v$ appears in the text $n$ published at time $t$. If it does, the element takes the value of one and zero otherwise. Hence, a specific text published at time $t$ can be represented by the row vector $\bm{c}_{n,t}$ (of size $1 \times V$). 

Typically, the corpus consists of a vast collection of texts, and our objective is to select texts for further analysis. For instance, we could focus on texts related to the U.S. economy from a collection of news articles published by various newspapers. A simple way to proceed is by using a keyword-based (\ie, token-based in our nomenclature) approach, which we formalize below.

We employ a selection matrix $\bm{\Omega}$ of size $V \times K$, where each column vector $\bm{\omega}^k$ (of size $V \times 1$) corresponds to a selection rule. The binary selection vector $\bm{\kappa}_t$ of size $N_t$ contains elements $\kappa_{n,t}$, which are defined through the selection function $f_\kappa(\cdot)$ as:
\begin{equation}\label{eq:filter}
\kappa_{n,t}  \equiv  f_\kappa(\bm{\Omega}, \bm{c}_{n,t}) \equiv I\bigg[\sum_{k=1}^{K} I\big[\bm{c}_{n,t} \bm{\omega}^k \ge 0\big] = \sum_{k = 1}^{K}I\big[\sum_{v = 1}^{V}\bm{i}_V \bm{\omega}^k > 0\big] \bigg]\,,
\end{equation}
where $I[\cdot]$ is an indicator function that takes a value of one if the condition inside the parenthese is true and zero otherwise.

In \eqref{eq:filter}, $\sum_{k=1}^{K} I\big[\bm{c}_{n,t} \bm{\omega}^k \ge 0\big]$, counts the number of unique active tokens (\ie, non-zero value) for dimension $k$ of the selection matrix $\bm{\Omega}$ (represented by $\bm{\omega}^k$) that appear in the text $n$ at time $t$ (represented by row vector $\bm{c}_{n,t}$). If this count is greater than zero for all dimensions $k=1,\ldots,K$, each element of the sum takes the value of one, and the summation takes the value $K$. The second part,  $\sum_{k = 1}^{K}I\big[\sum_{v = 1}^{V}\bm{i}_V \bm{\omega}^k > 0\big]$, where $\bm{i}_V$ is a row vector of lenght $V$ where each element is equal to one, counts the number of active tokens within each dimension of the selection matrix $\bm{\Omega}$. If this count is greater than zero for all dimensions, each element of the sum takes value one, and the sum is $K$. Thus, if the text $n$ published at time $t$ contains at least one active token within each non-zero column $k$ of $\bm{\Omega}$, the selection condition is satisfied, and $\kappa_{n,t}$ is equal to one.

The well-known Economic Policy Uncertainty (EPU) index of \citet{BakerEtAl2016} relies on this selection function. The EPU selection condition is defined by a vocabulary of size 10 with three selection dimensions ($V = 10$ and $K = 3$) representing economic, policy, and uncertainty-related tokens. Using our notation, the selection matrix $\bm{\Omega}$ (of size $10 \times 3$) of the EPU index is:
$$
\bm{\Omega}_{\text{EPU}} \equiv \kbordermatrix{
& \text{Economy} & \text{Policy} & \text{Uncertainty} \\
\text{economic} & 1 & 0 & 0 \\
\text{economy} & 1 & 0 & 0 \\
\text{congress} & 0 & 1 &0 \\
\text{deficit} & 0 & 1 & 0\\
\text{federal\_reserve} & 0 & 1 & 0\\
\text{legislation} & 0 & 1 & 0\\
\text{regulation} & 0 & 1 & 0\\
\text{white\_house} & 0 & 1 & 0\\
\text{uncertain} & 0 & 0 & 1 \\
\text{uncertainty} & 0 &  0 & 1 \\
} \,.$$
The EPU selection matrix contains 2, 6, and 2 active tokens in the first, second, and third dimension, respectively. In that case, the second terms of \eqref{eq:filter}: $\sum_{k = 1}^{K}I\big[\sum_{v = 1}^{V}\bm{i}_V \bm{\omega}^k_{\text{EPU}} > 0\big] = 3$.
For a text $n$ published at time $t$ containing the tokens ``economic,'' ``congress,'' and ``uncertainty,'' the first terms of  \eqref{eq:filter}: $\sum_{k=1}^{K} I\big[\bm{c}_{n,t} \bm{\omega}^k_{\text{EPU}} \ge 0\big] = 3$. Given the equality of both terms, for this article, $\kappa_{n,t}  = 1$. Alternatively, a text $n$ published at time $t$ containing two times the token ``economic'' and one time the token ``congress,'' the first terms of  \eqref{eq:filter}, $\sum_{k=1}^{K} I\big[\bm{c}_{n,t} \bm{\omega}^k_{\text{EPU}} \ge 0\big] = 2$, and thus $\kappa_{n,t}  = 0$ as both the left and right terms of \eqref{eq:filter} are not equal.

Several news-media-based indices follow similar selection rules: the Climate Policy Uncertainty index of \citet{gavriilidis2021measuring}, the Equity Market Volatility index of \citet{baker2019policy}, the Monetary Policy Uncertainty index of \citet{husted2020monetary}, and the Trade Policy Uncertainty index of \citet{handley2022trade}.

The selection function in \eqref{eq:filter} does not, however, take into account more complex selection mechanisms, such as the addition of a proximity condition between the tokens of two or more dimensions of the selection matrix $\bm{\Omega}$, as the one used to construct the Geopolitical Risk index of \citet{caldara2022measuring}. For instance, in the case of the EPU index, one might require at least one token from the ``Economy'' dimension detected in the neighborhood (\eg, within a two-token distance) of one of the ``Policy'' tokens in the same news article. Such additional complexity in the selection rule could be integrated using the information from a token distance matrix similar to the one used in \citet{ardia2021century}. To keep the exposition simple, we do not consider this type of variation of the selection function $f_\kappa(\cdot)$ in this paper. 

\subsection{Downstream Transformation} 

After identifying relevant texts in the corpus with the selection function $f_\kappa(\cdot)$, we typically transform and aggregate the selected texts into a quantitative time-series measure. We present below attention-based and context-based time-series measures.

An attention measure aims to measure the level of importance attributed to the selected texts across all texts in the corpus at a given time. For instance, it could be to measure the level of importance given to the topic of climate change by the news media \citep[see, \eg,][]{ArdiaEtAl2023}. Given our prior notation and definition, a typically used measure of attention is: 
\begin{equation}\label{eq:attention}
f_{\text{att}}( \bm{\Omega}, \bm{C}_t) \equiv \frac{1}{N_t}\sum_{n = 1}^{N_t} f_\kappa(\bm{\Omega}, \bm{c}_{n,t})\,.
\end{equation}

At each time $t$, we measure how many texts are selected and normalize this number by the total number of texts in the corpus at that time. If the corpus is composed of several sources (\eg, various newspapers), it may be relevant to standardize each source before the aggregation in \eqref{eq:attention}, as in  \citet{BakerEtAl2016}.

An alternative form of transformation is the content transformation, which requires more information than the output of the selection step. In particular, the content of selected texts is processed to derive a ``score,'' such as polarity or sentiment \citep[see, \eg][]{algaba2020econometrics}. The content transformation function can be written as:
\begin{equation}\label{eq:sentiment}
f_{\text{con}}(\bm{\Omega}, \bm{C}_t, \bm{\zeta}) \equiv \frac{1}{\sum_{n = 1}^{N_t} f_\kappa( \bm{\Omega},\bm{c}_{n,t})}\sum_{n = 1}^{N_t} 
\left( f_\kappa(\bm{\Omega},\bm{c}_{n,t})  \sum_{v=1}^{V} c_{n,v,t} \,  \zeta_v \right) \,,
\end{equation}
where $\zeta_v$ are token weights (integer or real numbers) used to score the selected texts. The average score of selected texts at time $t$ is then used to obtain an index. Token weights can be measured from manually-composed lexicons \citep[\eg,][]{loughran2014measuring} or in a data-driven way \citep[\eg,][]{jegadeesh2013word,manela2017news,kelly2021text,ardia2022media}. Much of the literature has focused on estimating $\bm{\zeta} \equiv (\zeta_1,\ldots,\zeta_V)$ given a predefined selection of texts. 

A practical advantage of employing the attention transformation over the content transformation lies in its reduced reliance on processing the entire corpus of texts through an additional layer of models to derive $\bm{\zeta}$ \citep[see, \eg,][]{algaba2020econometrics}. For instance, when utilizing a news articles aggregator that consolidates the desired news sources, the data required for computing attention solely invoke queries derived from a selection matrix (to determine the count of selected articles) and statistical information on the number of published news articles from the desired sources. This streamlined approach facilitates the timely computation and updating of the attention index. See, for instance, \url{https://www.policyuncertainty.com/}, where multiple indices similar to the EPU are updated monthly.

When using a data-driven method for content transformation, the joint estimation of the selection matrix $\bm{\Omega}$, and content transformation weights $\bm{\zeta}$, is preferable, albeit computationally intensive. Because of the added complexity of the content transformation, which is a challenge in this study, to isolate the effect of news selection from content transformation, we focus on the attention transformation measure.

\subsection{Objective-Based Tokens Optimization}
\label{opt_gamma}

The objective of deriving information in texts by a selection function and a transformation process is to capture a (contemporaneous or predictive) relation between that information and a variable of interest $y_t$ $(t=1,\ldots,T)$. A selection matrix is typically based on a subjective but guided assessment of the tokens necessary to capture the information we want to extract from the texts. Formally, given a  relevant transformation function $f(\bm{\Omega}, \bm{C}_t)$, the aim is to 
minimize: 
\begin{equation}\label{eq:optimization_base}
\min_{\alpha,\beta} \frac{1}{T}\sum_{t = 1}^{T}\left( y_t - \big(\alpha + \beta f(\bm{\Omega}, \bm{C}_{t-h})\big) \right)^2\,,
\end{equation}
with $h\geq0$, which is a simple linear regression problem.\footnote{While we present here the case of a linear relationship between $y_t$ and $f(\bm{\Omega}, \bm{C}_{t-h})$, more general specifications would work just as well, including multiple linear regression model by adding additional explanatory variables or by not assuming a linear relationship. To ease the presentation, we focus on a simple linear relationship without  additional explanatory variables.}  Once estimated, one would typically make inference about parameter $\beta$ or test whether the  (out-of-sample) root-mean-squared error is significantly lower than the root-mean-squared error of a model where $\beta$ is constrained to be zero (\ie, nested model); see \citet{ArdiaEtAl2019}. For instance, several studies analyze the relationship between the Economic Policy Uncertainty and economic variables using the following regression framework:\footnote{At the time of this writing, for instance, \citet{BakerEtAl2016}, the research paper introducing the EPU index, has received over 9,000 citations according to Google Scholar. While many of those citations do not use the index per se, many, however, analyze how the EPU explains or interacts with other macroeconomics and financial variables.} 
\begin{equation}
\min_{\alpha,\beta} \frac{1}{T}\sum_{t = 1}^{T}\left( y_t - \big(\alpha + \beta f_{\text{att}}(\bm{\Omega_{\text{EPU}}}, \bm{C}_{t-h})\big)\right)^2\,.
\end{equation}

One limitation of this approach is related to the fact that $\bm{\Omega}_\text{EPU}$ is given. This may be reasonable in some applications but suboptimal for others, particularly when it comes to nowcasting and forecasting. We thus introduce the following optimization problem:
\begin{equation}\label{eq:optimization} %\label{eq:optimization_extended5}
\min_{\alpha,\beta,\bm{\Omega}} 
\frac{1}{T}\sum_{t = 1}^{T}\left(y_t - \big(\alpha + \beta f_{\text{att}}(\bm{\Omega}, \bm{C}_{t-h}) \big)\right)^2
+ \lambda_1 \sum_{v = 1}^{V}I[\bm{\omega}_v \bm{i}'_K> 1] + \lambda_2  I[\beta > 0] + \lambda_3  I[\beta \leq 0]\,.
\end{equation}
where $\bm{\omega}_v$ denotes the $v$th row of $\bm\Omega$. In optimization problem \eqref{eq:optimization}, we assume an unrestrictive and large vocabulary and optimize active tokens selection in addition to the regression parameters. We use three penalty terms to control the optimization process. We standardize $y_t$ to avoid scale-dependent penalty parameters.

First, we introduce a penalty term $\lambda_1 \sum_{v = 1}^{V}I[ \bm{\omega}_v \bm{i}'_K > 1]$, which aims to avoid the activation of the same token across multiple dimensions in the selection matrix. A higher value of $\lambda_1$ enforces a constraint that restricts each dimension from overlapping. This constraint can be beneficial when assigning specific topics to each dimension. However, it also limits the flexibility of the selection matrix. Consider the following two three-dimensional selection matrices:
$$
\kbordermatrix{
& \text{1} & \text{2} & \text{3} \\
\text{economic} & 1 & 1 & 1 \\
\text{stock\_market} & 1 & 0& 0\\
\text{federal\_reserve} & 0 & 1 &0 \\
\text{crisis} & 0 & 0 & 1 \\
} \quad \quad
\kbordermatrix{
& \text{1} & \text{2} & \text{3} \\
\text{economic} & 1 & 0 & 0 \\
\text{stock\_market} & 1 & 0& 0\\
\text{federal\_reserve} & 0 & 1 &0 \\
\text{crisis} & 0 & 0 & 1 \\
} 
$$

With the first selection matrix, a text containing the token ``economic'' would be selected, while it would not necessary with the second, as in this case, a text needs to include tokens from the other dimensions, distinct from ``economic.'' Therefore, a lower value of $\lambda_1$ allows for the inclusion of selection conditions with lower dimensions than the one of the selection matrix, meaning, for instance, a text may fulfill the requirement with a single distinct token instead of three distinct tokens (when $K = 3$). Setting $\lambda_1 = 0$ provides an opportunity for a higher-dimensional selection matrix to mimic a lower-dimensional one, as shown by these two equivalent solutions:
\[
\bm{\Omega} = \kbordermatrix{
& \text{1} & \text{2} & \text{3} \\
\text{economic} & 1 & 1 & 1 \\
}=
\kbordermatrix{
& \text{1} \\
\text{economic} & 1  \\
}
\]

Increasing flexibility with $\lambda_1 = 0$ increases complexity, which may lead to challenges in the selection matrix's out-of-sample performance. A higher value of $\lambda_1$ penalizes the score of the left selection matrix and, therefore, leads to a narrower set of potential solutions. 

The two remaining penalty terms, $\lambda_2 I[\beta > 0]$ and $\lambda_3 I[\beta \leq 0]$, serve to establish the intended association between the text-based index and the variable of interest $y_t$. When a negative relationship is sought, the condition $\lambda_2 > \lambda_3$ is applied. Consequently, this allows for controlling a selection condition that selects texts either positively or negatively correlated with the variable of interest. It is crucial to choose the sign of beta as selection matrices can be derived in a manner that selects texts associated negatively with the variable of interest and another set that is positively associated with it. %For instance, in the context of stock market returns, one selection matrix might consider texts indicative of positive market sentiment ($\beta > 0$), while another could select texts reflecting negative sentiment regarding the stock market ($\beta < 0$). Without this precondition, the algorithm will converge toward the sign that provides greater explanatory power, never generating a selection matrix selecting texts from the other sign, even though they might be as relevant.

\section{Optimization Strategy}
\label{sec:optimization} 

Regression \eqref{eq:optimization_base} is trivial to estimate as $\bm{\Omega}$ is fixed and thus not part of the estimation process. On the contrary, the minimization problem \eqref{eq:optimization} is complex as $f(\bm{\Omega}, \bm{C}_t)$ is a non-linear function of $\bm{\Omega}$, which is now considered as a parameter. Below, we present our strategy based on a 
genetic algorithm to perform the minimization. We first define the crossover and mutation operators specifically designed for our problem. 

\subsection{Crossover and Mutation Operators}

Because of the distinctive characteristics of our optimization problem, we introduce specific crossover and mutation operators. To illustrate these operators, we consider a specific scenario wherein our optimization objective is to replicate the EPU index while imposing a constraint that restricts the number of active tokens to five out of a maximum of $V\!K$ (\ie, when all elements in the matrix $\bm{\Omega}$ are equal to one). %We present a ``best-case'' scenario for each operator to elucidate the effectiveness and rationale underlying these operators, demonstrating their functionalities and objectives. \dave{XXX pas clair}

\subsubsection{Token Crossover Operator}

We begin with the ``token crossover operator.'' This operator takes two parent solutions and generates two offspring solutions by exchanging a single token between the parents while maintaining the dimensionality of the replaced token. Consider the following two parents: 
$$
\kbordermatrix{
\text{Parent 1:} & 1 & 2 & 3 \\
\CB{\text{economic}} & 1 & 0 & 0 \\
\text{legislation} & 0 & 1 & 0 \\
\text{congress} & 0 & 1 & 0 \\
\text{house} & 0 & 0 & 1\\
\text{risk} & 0 & 0 & 1 \\
\vdots & \vdots & \vdots &\vdots
} \quad\quad
\kbordermatrix{
\text{Parent 2:} & 1 & 2 & 3 \\
\text{economy} & 1 & 0 & 0 \\
\CBX{\text{financial\_crisis}} &0 & 1 & 0 \\
\text{congress} & 0 & 1 & 0 \\
\text{house} & 0 & 0 & 1\\
\text{risk} & 0 & 0 & 1 \\
\vdots & \vdots & \vdots &\vdots
}
$$

The token crossover operator selects two tokens (in gray) and generates 
two offspring solutions as follows:
$$
\kbordermatrix{
\text{Child 1} & 1 & 2 & 3 \\
\CBX{\text{financial\_crisis}} & 1 & 0 & 0 \\
\text{legislation} & 0 & 1 & 0 \\
\text{congress} & 0 & 1 & 0 \\
\text{house} & 0 & 0 & 1\\
\text{risk} & 0 &  0 & 1 \\
\vdots  & \vdots & \vdots  &\vdots
} \quad \quad
\kbordermatrix{
\text{Child 2} & 1 & 2 & 3 \\
\text{economy} & 1 & 0 & 0 \\
\CB{\text{economic}} &0 &  1 & 0 \\
\text{congress} & 0 & 1 & 0 \\
\text{house} & 0 & 0 & 1\\
\text{risk} & 0 &  0 & 1 \\
\vdots  & \vdots & \vdots  &\vdots
}
$$
It is important to note that this crossover operator maintains a fixed number of active tokens within each dimension and operates solely on already activated tokens. In contrast, a regular crossover operator would not guarantee that each dimension remains the same size or even has active tokens. Additionally, owing to the high sparsity of our optimization problem (where only five tokens are active out of a potential $V\!K$ points), a regular crossover operator could perform non-altering operations by exchanging slices of non-active tokens (\ie, tokens with zero values), consequently reducing the efficiency of the search algorithm.

\subsubsection{Mutation Operators}

Next, we focus on three mutation operators, each serving a distinct purpose. 

\paragraph{Switch Mutation Operator} The ``switch mutation'' operator allows an active token to change its dimension:
$$
\kbordermatrix{
\text{Parent} & 1 & 2 & 3  \ \\
\text{economy} & 1 & 0 & 0 \\
\text{economic} &0 & \CB{1} & 0 \\
\text{congress} & 0 & 1 & 0 \\
\text{house} & 0 & 0 & 1\\
\text{risk} & 0 &  0 & 1 \\
\vdots  & \vdots & \vdots  &\vdots
} \quad \quad
\kbordermatrix{
\text{Child}	 & 1 & 2 & 3 \\
\text{economy} & 1 & 0 & 0 \\
\text{economic} &\CB{1} &  0 & 0 \\
\text{congress} & 0 & 1 & 0 \\
\text{house} & 0 & 0 & 1\\
\text{risk} & 0 &  0 & 1 \\
\vdots  & \vdots & \vdots  &\vdots
}
$$

This operator allows for the correction of potential dimensional misalignment. It is worth noting that in the example above, the set of texts selected by the parent, while not optimal, may be highly correlated with the set of texts of the child, which is optimal. This operation facilitates a more efficient search for misalignment compared to a random search. Similar to other operators, this mutation operator only applies to active tokens. However, it cannot change the dimensions of a token if it is the only token within its dimension, ensuring that each dimension has at least one active token.

\paragraph{N-Gram Mutation Operator} 

Next, we introduce the ``n-gram mutation'' operator, which involves transforming one of the selected tokens into an n-gram containing that token:
$$
\kbordermatrix{
\text{Parent} & 1 & 2 & 3  \\
\text{economy} & 1 & 0 & 0 \\
\text{economic} & 1 &  0 & 0 \\
\text{congress} & 0 & 1 & 0 \\
\CB{\text{house}} & 0 & 0 & 1\\
\text{risk} & 0 &  0 & 1 \\
\vdots  & \vdots & \vdots  &\vdots
}
\quad \quad
\kbordermatrix{
\text{Child}	 & 1 & 2 & 3 \\
\text{economy} & 1 & 0 & 0 \\
\text{economic} & 1 &  0 & 0 \\
\text{congress} & 0 & 1 & 0 \\
\CB{\text{white\_house}} & 0 & 0 & 1\\
\text{risk} & 0 &  0 & 1 \\
\vdots  & \vdots & \vdots  &\vdots
}
$$

To comprehend this operator, we must recognize that the texts containing an n-gram always include the texts containing its components. For instance, the number of texts containing the token ``stock'' is equal to or larger than the number of texts containing the bi-gram ``stock\_market.'' As such, if the algorithm selects the component of an n-gram, there is a high likelihood that using an n-gram containing that component would be more optimal. The n-gram mutation operator increases the possibility of testing an n-gram compared to testing any other token. Similar to the crossover operator, the n-gram mutation only applies to active tokens, but it also applies solely to tokens that are part of n-grams within the vocabulary.

\paragraph{Transform Mutation Operator} Finally, we have the ``transform mutation'' operator, which changes a token from a specific dimension to another token in that same dimension:
$$
\kbordermatrix{
\text{Parent} & 1 & 2 & 3  \\
\text{economy} & 1 & 0 & 0 \\
\text{economic} & 1 &  0 & 0 \\
\text{congress} & 0 & 1 & 0 \\
\text{white\_house} & 0 & 0 & 1\\
\CB{\text{risk}} & 0 &  0 & 1 \\
\vdots  & \vdots & \vdots  &\vdots
} \quad \quad
\kbordermatrix{
\text{Child} & 1 & 2 & 3\\
\text{economy} & 1 & 0 & 0 \\
\text{economic} & 1 &  0 & 0 \\
\text{congress} & 0 & 1 & 0 \\
\text{white\_house} & 0 & 0 & 1\\
\CB{\text{uncertainty}} & 0 &  0 & 1 \\
\vdots  & \vdots & \vdots  &\vdots
}
$$
The transform mutation is the only operator that activates new tokens (not n-grams) that have not already been activated within the population. This operator thus introduces new tokens within the population by replacing an active token from a candidate selection matrix with a new one. The transform mutation ensures that the number of active tokens within each dimension remains the same. This operator is crucial in controlling the number of active tokens while allowing the introduction of new active tokens. A standard mutation operator that flips one bit would not work in this case. Consider a vocabulary of size 4,800 with three dimensions, leading to $4,\!800 \times 3 = 14,\!400$  potential number of active points. If a parent has ten active tokens out of 14,\!400, the standard mutation operator has a $\frac{14,\!400-10}{14,\!400})= 99.93\%$ chance of generating a child with 11 active tokens and a $0.07\%$ chance of generating a child with nine active tokens. Over a large number of iterations, this leads to a search space with a large number of active tokens, ultimately leading to overfitting. 

By design, many of these operators only apply to tokens that are already active within the corpus to ensure an efficient search. Due to the extreme sparsity of our optimization problem, without these operators, the likelihood of performing non-altering operations is extremely high. We also note that these operations ensure that the number of active tokens in the child will always be the same as the number of active tokens in the parent. This is crucial for our search strategy, which is presented below.

\subsection{Calibration}

The calibration strategy is centered around (i) efficient search and (ii) avoiding overfitting. First, we define a training and a validation window. We start the optimization process by defining the initial population of $q = 1, \dots, Q$ candidate selection matrices $\bm{\Omega}_q$ of size $V \times K$. Each member of the initial population starts with $K$ active tokens (non-zero value in the selection matrix), where $K$ is the number of dimensions of the selection matrix. Each dimension is set to have one active token. For instance, for $K = 3$, these could be members of the initial population (displaying only active tokens):
\[
\kbordermatrix{
& 1 & 2 & 3  \ \\
\text{democrats} & 1 & 0 & 0 \\
\text{european\_union} & 0 & 1 & 0 \\
\text{policy} & 0 & 0 &1 \\
\vdots  & \vdots & \vdots  &\vdots
} \quad\quad
\kbordermatrix{
& 1 & 2 & 3  \ \\
\text{religion} & 1 & 0 & 0 \\
\text{market} & 0 & 1 & 0 \\
\text{presidency} & 0 & 0 &1 \\
\vdots  & \vdots & \vdots  &\vdots
}
\]
while these would not:

\[
\kbordermatrix{
& 1 & 2 & 3  \ \\
\text{democrats} & 1 & 0 & 0 \\
\text{european\_union} & 0 & 0 & 1 \\
\text{policy} & 0 & 0 &1 \\
\vdots  & \vdots & \vdots  &\vdots
} \quad\quad
\kbordermatrix{
& 1 & 2 & 3  \ \\
\text{religion} & 1 & 0 & 0 \\
\text{market} & 0 & 1 & 0 \\
\text{presidency} & 0 & 0 &1 \\
\text{coffee} & 0 & 0 &1 \\
\vdots  & \vdots & \vdots  &\vdots
}
\]

Inpired from \citet{Scrucca2013,Scrucca2017}, we perform genetic optimization as follows:

\begin{enumerate}

\item Begin an epoch with an initial population. For the first epoch, this is randomly initialized, while for subsequent epochs, the initial population depends on the results from the previous epoch. Each epoch contains $H$ iterations along these steps:

\begin{enumerate}
	
	\item For each candidate $q=1,\ldots,Q$ in the population, compute $f(\bm{\Omega}_q, \bm{C}_{t-h})$. 
	
	\item For each candidate $q=1,\ldots,Q$ in the population, estimate parameters $\alpha$ and $\beta$ by ordinary least squares on the training data with the computed $f(\bm{\Omega_q}, \bm{C}_{t-h})$ as input. 

	\item For each candidate $q=1,\ldots,Q$ in the population, compute the objective function \eqref{eq:optimization} on the training and validation sets using the estimated parameters from step 2. 

	\item Elitism step. Store the 5\% lowest objective value chromosome in the training window as well as the best solution. For the best solution, store its validation window objective value. 5\% is a common choice for  elitism steps in genetic algorithms. 
	
	\item Tournament selection. Select at random three members of the population, then select the one with the best objective function. Do this $Q$ times to generate an interim population. %Generate an interim population by selecting $Q$ selection matrices using tournament selection (of size 3).\keblu{}

	\item Token crossover: For pairs of two randomly chosen selection matrices in the interim population, perform a token crossover operation with probability $S$ to update the interim population, where $S = \frac{Q_{\text{unique}}}{Q}$ is the ratio of the unique selection matrix in the interim population over the population size. Since doing crossover on two populations that are exacty the same does not do anything, the probability of doing a crossover operation scale. If all matrices are unique ($S = 1$), we would perform crossover on all population members (100\% probability); if all matrices are the same, $S \approx 0$, no crossover is performed.

	\item Switch mutation. For each selection matrix in the interim population, perform a switch mutation operation with 5\% probability to update the interim population. 5\% is chosen arbitrarily. 

	\item N-grams mutation. For each selection matrix in the interim population,  perform n-grams mutation operation with 5\% probability to update the interim population. 5\% is chosen arbitrarily. 

	\item Transform mutation. For each selection matrix in the interim population, perform a transform mutation with $1-S$\% probability to update the interim population. The transform mutation changes an active word for another non-active word. If all matrices are the same (S $\approx $ 0), all members of the interim population will have a transform mutation operator applied to them. If all matrices are unique, no transform mutation will be applied. This overall strikes a balance between the crossover and transform mutation operator, thus leveraging crossover when appropriate and doing the equivalent of pure random search when there is no value to performing the crossover operator.

	\item Elitism step. Replace 5\% instance of the interim population with the 5\% lowest objective value solution stored in step (d). We keep the best members of the populations before crossover and mutation. This strategy usually speeds up the convergence of the algorithm.

	\item Start a new iteration with the new population until we reach the last iteration.
\end{enumerate}

\item From $H$ best solutions, each one stored at step (d), Select the one with the lowest validation window objective value.

\item Refine the vocabulary for the next epoch. To achieve this, we employ word embeddings to reduce the vocabulary size while retaining information from the active tokens of the best solution in the preceding step. A word embedding space maps tokens into high-dimensional vectors, where similar tokens exhibit shorter distances between their vectors. First, we compute the average token vector for each dimension of the best solution in step 2. Then, for each dimension, we calculate the cosine similarity between the average token embedding of that dimension and all the tokens in the original vocabulary of size $V$. We retain tokens with a cosine similarity larger than 0.2. The set of unique, similar tokens across all dimensions, constitutes our new vocabulary. Effectively, this step narrows down the search space to a more relevant vocabulary, considering that the optimally selected tokens are indicative of the relevant tokens concerning the dependent variable.

\item Subsequently, we form a new population of size $Q$ using the new vocabulary, where each member starts with $K+1$ active tokens, one more than the old population. Additionally, we insert the optimal solution from step 2  within the new population. If we find a better solution with more active tokens, it will disappear from the population. 

\item Start a new epoch until we achieve the maximum number of desired active tokens (from $K$ up to 15 in our application).

\end{enumerate}

The solution with the lowest objective value from the validation window from step 2 of the last epoch is considered as the interim optimal solution. To obtain the final solution, we perform a pruning step. This step aims to eliminate tokens that may have adversely affected the model fit but were added due to the random nature of the optimization process (\eg, uninformative tokens added before informative ones). To accomplish this, we test variations of the optimal selection matrix by deactivating one or several active tokens, encompassing all variations involving only deactivations. For instance, if the solution with the lowest objective value from the validation window has 12 active tokens, we test $2^{12}= 4,\!096$ potential new solutions (ranging from all tokens being active to all tokens being deactivated). %The new optimal solution could only contain ten active tokens by deactivating two active words, and we would select it as the final solution based on the lowest objective value from the combined training and validation window among all newly tested candidates.

\section{Empirical Applications}
\label{sec:empirical} 

In this section, we present two empirical applications of our methodology. First, we validate our approach with the EPU index. Second, we apply our optimization process to (i) tracking the VIX and (ii) nowcasting inflation expectations. We use the latter two usecases as news-based indices have been used in previous studies in relation to those variables, and as such, we can test our method against benchmarks. Beforehand, we describe the corpus of text used throughout this section and the data processing steps.

\subsection{Textual Data, Corpus Processing, and Vocabulary}

Our corpus consists of all news articles published in the printed version of the Wall Street Journal from January 2000 to August 2021. In total, the corpus is composed of 763,542 news articles. To derive candidate tokens for the vocabulary describing the corpus, we proceed as in \citet{ArdiaEtAl202x}:
\begin{enumerate}
\item To normalize the news articles, we lowercase and lemmatize each word into its root form.

\item We then detect and combine collocation in our corpus. For instance, the compound words ``interest rate'' or ``exchange rate'' provide more information about the content of an article than if these words were not considered a combination. We rely on two methods to detect collocations in our documents: (i) The RAKE (Rapid Automatic Keyword Extraction) algorithm \citep{RoseEtAl2010} and (ii) the process described in \citet[Section IV.B]{HansenEtAl2018}.\footnote{Specifically, RAKE assigns a score to each candidate keyword based on the frequency of occurrence of the keyword in the text, as well as the frequency of occurrence of each of its constituent words in the text. The algorithm then identifies sets of adjacent keywords with high scores, indicating that they are likely collocations. The approach in \citet[Section IV.B]{HansenEtAl2018} looks at part of speech patterns within the text. Following \citet{JustesonKatz1995} these patterns are: adjective-noun, noun-noun, adjective-adjective-noun, adjective-noun-noun, noun-adjective-noun, noun-noun-noun, and noun-preposition-noun. We find the part of speech of each word using the UDPipe methodology implemented in the \R package \pkg{udpipe} \citep{WijffelsEtAl2021}.} For each method, we select the collocations that occur at least 100 times for bigrams and 50 times for trigrams.  Finally, we concatenate the individual tokens of the collocation in each text of our corpus (\eg, ``interest rate'' becomes ``interest\_rate''). 

\item We then build a matrix representation of the text where each column is a token, and each row is a news article such that each element represents the number of times a token is observed for a corresponding news article. We eliminate tokens that appear in less (more) than 1\% (20\%) of the news articles. 

\item Finally, we manually verify each token and remove non-informative tokens such as dates or tokens with numbers. We also remove any token related to specific events, persons, locations, companies, or products. Overall, this results in a vocabulary of size $V = 2,\!261$.
\end{enumerate}

We use the pre-trained continuous bag-of-word embedding model of \citet{rahimikia2021realised} available at \url{https://fintext.ai} to retrieve the token vector for each of the 2,261 tokens. The model is compiled from 15 years of business news archives and, as a domain-specific model, is more appropriate to capture the relationship between tokens than a pre-trained general-domain word embedding model such as  Google Word2Vec. For collocations, we take the average of the token vector across the individual collocation tokens. These token vector representations of the 2,261 tokens will be used to narrow down the vocabulary after each epoch.  

\subsection{Validation of our Algorithm With the Economic Policy Uncertainty}

Our first application aims to validate our approach by determining if we can recover the EPU selection matrix. Specifically, we will perform the following optimization problem:

\begin{align}\label{eq:EPU}
\begin{split}
(\hat \alpha, \hat \beta, \hat{\bm{\Omega}}) \equiv \argmin_{\alpha,\beta,\bm{\Omega}} \frac{1}{T}\sum_{t = 1}^{T} \left( f_{\text{att}}(\bm{\Omega_{\text{EPU}}}, \bm{C}_t) - \big(\alpha 
+ \beta f_{\text{att}}(\bm{\Omega}, \bm{C}_t)\big)\right)^2 \\
+ \lambda_1 \sum_{v = 1}^{V} I[ \bm{i}_K \bm{\omega}_v > 1] + \lambda_2  I[\beta > 0] + \lambda_3  I[\beta \leq 0]\,,
\end{split}
\end{align}	

and see if $\hat{\bm{\Omega}} = \bm{\Omega_{\text{EPU}}}$. With $V = 2,\!261$ and $K = 3$, the number of selection matrices is $2^{2261\times 3}$. Convergence toward the true EPU selection matrix would thus indicate massive improvement compared to a naive random search.

The setup for this experiment is as follows. We set the size of the selection matrices population to $Q = 300$, the number of iterations per epoch to $H = 200$, and the number of epochs to 15. The training window is composed of monthly data from January 2000 to December 2010 (132 observations), and the validation window ranges from January 2011 to December 2014 (48 observations). The remaining data from January 2015 to August 2021 is our test window (80 observations). We set $\lambda_1 = 0.25$, $\lambda_2 = 0$, and $\lambda_3 = 1$, so that we force the algorithm to search for tokens that have a positive relationship to the dependent variable. We test our algorithm for selection matrices with dimensions $K = 1,2,3$. 

The optimal selection matrices obtained for each dimension are shown below:
\[
\hat{\bm{\Omega}}_{K = 1} \!=\!\!\! \!\!\kbordermatrix{
& 1 \ \\
\text{uncertainty} & 1 \\
\text{uncertain} & 1  \\
\text{loom} & 1  \\
\text{unknown} & 1  \\
\vdots  & \vdots
} \quad\!\!
\hat{\bm{\Omega}}_{K = 2} \!=\!\!\!\!\!\kbordermatrix{
& 1 & 2   \ \\
\text{uncertainty} & 1 & 0 \\
\text{uncertain} & 1 & 0  \\
\text{monetary} & 0 & 1  \\
\text{banking} & 0 & 1  \\
\text{regulate} & 0 & 1 \\
\text{economic\_growth} & 0 & 1  \\
\text{spending} & 0 & 1  \\
\text{overhaul} & 0 & 1  \\
\text{extend} & 0 & 1 \\
\text{necessary} & 0 & 1  \\
\text{reserve} & 0 & 1  \\
\text{democrats} & 0 & 1  \\
\vdots  & \vdots & \vdots
} \quad
\hat{\bm{\Omega}}_{K = 3} \!= \!\!\!\!\!\kbordermatrix{
& 1 & 2 & 3  \ \\
\text{uncertainty} & 1 & 0 & 0\\
\text{uncertain} & 1 & 0 & 0  \\
\text{deficit} & 0 & 1 & 0  \\
\text{congress} & 0 & 1 & 0  \\
\text{legislation} & 0 & 1  & 0\\
\text{regulation} & 0 & 1 & 0  \\
\text{federal\_reserve} & 0 & 1 & 0  \\
\text{white\_house} & 0 & 1 & 0  \\
\text{economic} & 0 & 0 & 1 \\
\text{economy} & 0 & 0 & 1  \\
\vdots  & \vdots & \vdots
}
\]

We note that $\hat{\bm{\Omega}}_{K = 3}$ is exactly the EPU selection matrix, thus confirming that our algorithm can converge to an exact solution even with such a large search space. 

In Table \ref{tab:epu}, we report the number of active tokens and the  training window, validation window, and test window root mean-squared-errors (RMSE) for each epoch, along with the optimal results after the pruning step. We notice that the reduction in all error measures is much less pronounced when the wrong number of dimensions of the selection matrix is used. %Also, a two-dimensional selection matrix better approximates a three-dimensional selection matrix than the approximation that a one-dimensional selection matrix would give. It is, however, hard to generalize the error rate (in the test set) when the wrong number of dimensions is used. In any case, we would use the error of the pruning step as a basis for selecting the number of dimensions, which would have resulted in the correct choice.
We also see that the pruning step was necessary for retrieving the exact EPU matrix when optimizing the three-dimensional selection matrix. Indeed, on the 13 epochs, the three-dimensional selection matrix had 11 active tokens, of which one was removed after pruning. Finally, we note that no active token is added for some epochs, but for subsequent epochs, more than one active token is added. This might be because a token is only more informative in the presence of another token and not by itself.

The computational time on a standalone computer (single Intel core i9 3.7GHz with 48 GB RAM) ranges from five hours and 25 minutes for $K=1$ to nine hours and 45 minutes for $K=3$. Overall, this validation exercise demonstrates that our optimization can recover pre-defined selection matrices within a reasonable amount of 
computational time. 

\insertfloat{Table \ref{tab:epu}}

\subsection{VIX and Inflation Expectations}

Next, we test our methodology for (i) tracking the VIX index and (ii) nowcasting inflation expectations. We use those two usecases because news articles were previously used to build text-indices to forecast these variables. Thus, we have natural benchmarks to compare the performance of our approach.

For the VIX, \citet{baker2019policy} construct the Equity Market Volatility (EMV) index, which, in our framework, can be defined with the following selection matrix:

\[
\bm{\Omega}_{\text{EMV}} \equiv \kbordermatrix{
& \text{Equity} & \text{Market} & \text{Volatility} \\
\text{economic} & 1 & 0 & 0 \\
\text{economy} & 1 & 0 & 0 \\
\text{financial} & 1 & 0 &0 \\
\text{stock\_market} & 0 & 1 & 0\\
\text{equity} & 0 & 1 & 0\\
\text{equities} & 0 & 1 & 0\\
\text{standard\_and\_poors} & 0 & 1 & 0\\
\text{s\&p} & 0 & 1 & 0\\
\text{volatility} & 0 & 0 & 1 \\
\text{volatile} & 0 &  0 & 1 \\
\text{uncertain} & 0 & 0 & 1 \\
\text{uncertainty} & 0 &  0 & 1 \\
\text{risk} & 0 & 0 & 1 \\
\text{risky} & 0 &  0 & 1 \\
}
\]

This news selection is then further analyzed and divided to quantify journalist perceptions about
the news items, developments, concerns, and anticipations that drive volatility in equity returns. By optimizing the selection matrix, we aim to derive an attention index that more closely matches the VIX  than the EMV does.\footnote{The VIX is retrieved from \url{https://fred.stlouisfed.org/series/VIXCLS}.} This is particularly useful as a better overarching index incorporates a more precise selection for selecting news having a positive relationship with market volatility, and thus better explains, in a quantitative manner (due to the availability of news data), the level of volatility. This could then be decomposed into a more narrow indicator as in \citet{baker2019policy}.

Regarding inflation expectations, \citet{pfajfar2013news} and \citet{marcellino2022demand} analyze the link between news about inflation and how households form expectations about inflation. In particular, \citet{marcellino2022demand}  conclude that media communication and agents' attention play an important role in aggregate inflation expectations. In our framework, their attention index can be defined as:\footnote{In particular \citet{marcellino2022demand}, search for instance of words containing ``inflat'' while \citet{pfajfar2013news} search for words whose root is ``inflation.'' Since our words are lemmatized, all those instances are included in the word ``inflation.''}
$$
\bm{\Omega}_{\text{INFL}} \equiv \kbordermatrix{
& \text{inflation}   \\
\text{inflation} & 1  \\
}
$$

For a measure of inflation expectations, we use data from the Michigan University Survey of Consumers; we consider the median expected price change for the next 12 months (MICH).\footnote{MICH is retrieved from \url{https://fred.stlouisfed.org/series/MICH}.} Similar to the VIX, we want to find a selection matrix that generates a news attention index that more closely matches inflation expectations. This, in turn, can be used to determine better what news are related to inflation expectations by analyzing the tokens in the selection matrix or by further analyzing the news selected by the selection matrix (for instance, via a topic model).

The time-series indices constructed with $\bm{\Omega}_{\text{EMV}}$ and $\bm{\Omega}_{\text{INFL}}$ will serve as benchmarks when evaluating the fit on the test window. For the optimization, we use a population of size $Q = 200$ and $H = 500$ iterations per epoch with a maximum of 15 tokens (and, as such 13-15 epochs depending on the starting number of filtering dimensions). We set $\lambda_1 = 0.25$ and optimize for tokens positively correlated with the target index ($\lambda_2 = 0$ and $\lambda_3 = 1$). We keep the solution among $K = 1,2,3$ with the lowest objective value from the combined training and validation windows. Finally, we benchmark the results against $f_{\text{att}}(\bm{\Omega_{\text{EMV}}}, \bm{C}_t)$ and $f_{\text{att}}(\bm{\Omega_{\text{INFL}}}, \bm{C}_t)$ for the VIX and inflation expectations, respectivelly. 

The best selection matrices obtained on the combined training and validation windows after pruning are shown below: 

\[
\hat{\bm{\Omega}}_{\text{EMV}} =\!\!\!\! \kbordermatrix{
& & \\
\text{economy } & 1& 0 \\
\text{stock\_market } & 1 & 0 \\
\text{crisis} & 1 & 0 \\
\text{volatility} & 1& 0  \\
\text{despite} & 1 & 0 \\
\text{club} & 1& 0  \\
\text{fear} & 0 & 1 \\
\text{plunge} & 0 & 1 \\
\text{industrial\_average} & 0 & 1 \\
\text{package} & 0  & 1\\
\text{summit} & 0& 1 \\
\text{slash} &0 & 1 \\
\text{antitrust} & 0 & 1 \\
\vdots  & \vdots
} \quad \quad
\hat{\bm{\Omega}}_{\text{INFL}} =\!\!\!\! \kbordermatrix{
& & \ \\
\text{food                  } & 1& 0 \\
\text{customer                   } & 1 & 0 \\
\text{frustration                   } & 1& 0  \\
\text{hardware} & 1 & 0 \\
\text{vote                       } & 1& 0  \\
\text{involved                     } & 1& 0  \\
\text{disruption                    } & 1 & 0 \\
\text{clothes } & 1 & 0 \\
\text{suspend                 } & 1 & 0 \\
\text{shortage                       } & 0 & 1 \\
\text{gasoline                 } & 0 & 1 \\
\text{inflation                 } & 0& 1  \\
\vdots  & \vdots
}
\]

A two-dimensional selection matrix generates the best performance in both cases. We see that most selected tokens by our optimization strategy have an intuitive relation with the target variable. This data-driven selection could be improved either (i) by augmenting the population size, the number of iterations, or the number of active tokens in the algorithm or (ii) by proceeding with a final human-supervised pruning step. 

Table \ref{tab:vix} reports the root mean squared error (RMSE) of linear regression models applied to tracking the VIX and inflation expectations using benchmarks and optimal indices across different time windows. As expected, the optimization process leads to an optimal index that exhibits lower RMSE for both the VIX and inflation expectations during the training and validation periods. The reduction in RMSE is substantial for the VIX (49\% in the training window and 39\% in the validation window) and inflation expectations (23\% in the training window and 27\% in the validation window). More importantly, the outperformance persists on the test window (data not observed during the selection matrix optimization process), reducing RMSE by 37\% for the VIX and 25\% for inflation expectations relative to the benchmarks.

\insertfloat{Table \ref{tab:vix}}

\begin{figure}[H]
\centering
\includegraphics[width=0.95\textwidth]{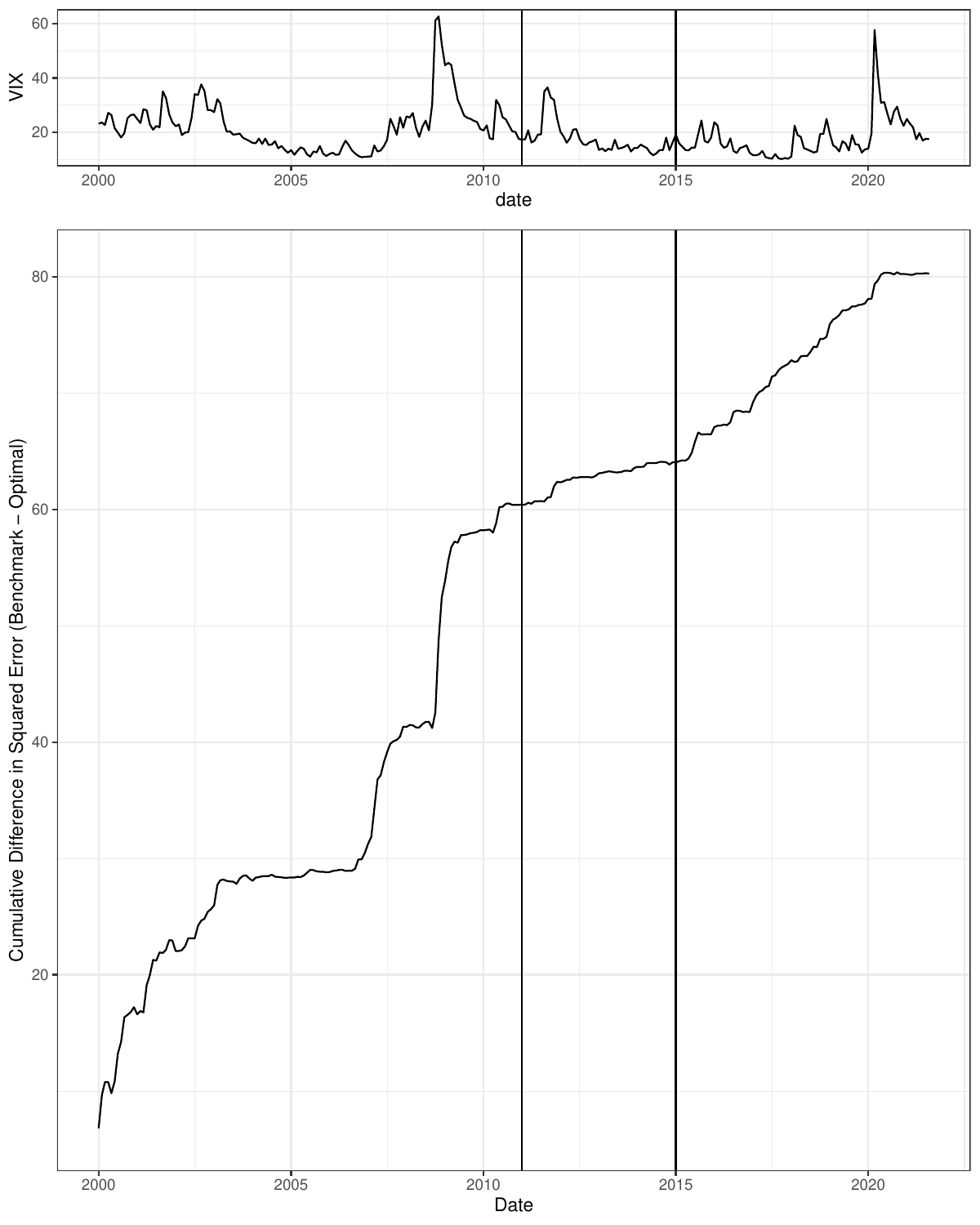}
\caption{\textbf{Cumulative Squared Error Difference -- VIX}\\
This figure shows the cumulative squared error difference for the results generated by the benchmark selection matrix minus the one generated by the optimal selection matrix for the VIX usecase. Vertical lines indicate the end of the training window and the validation window. }
\label{fig:VIX}
\end{figure}

\begin{figure}[H]
\centering
\includegraphics[width=0.95\textwidth]{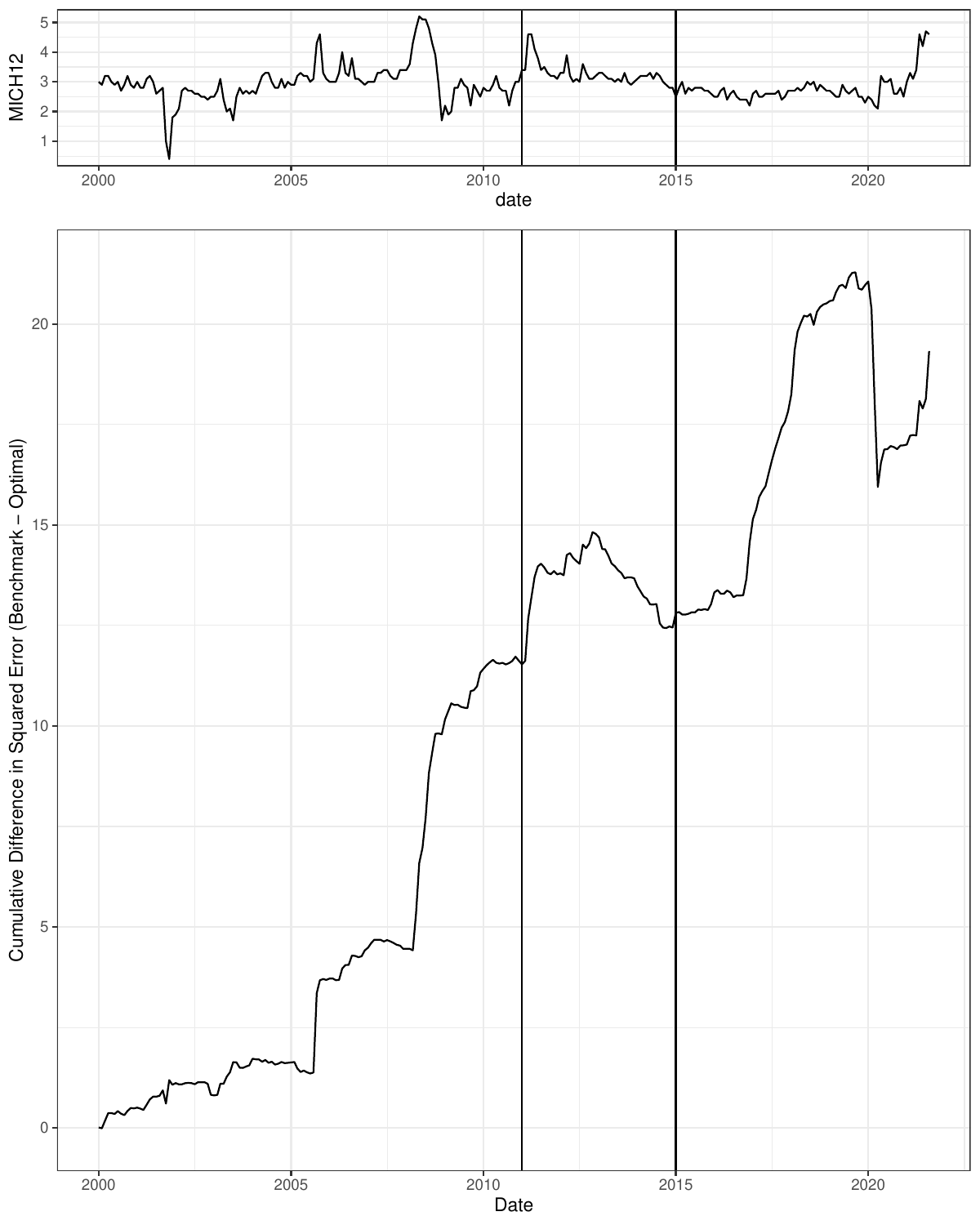}
\caption{\textbf{Cumulative Squared Error Difference -- MICH}\\
This figure shows the cumulative squared error difference for the results generated by the benchmark selection matrix minus the one generated by the optimal selection matrix for the MICH (twelve-month inflation expectation) usecase. Vertical lines indicate the end of the training window and the validation window.}
\label{fig:MICH}
\end{figure}

Further examination of the differences in performance is provided in Figures \ref{fig:VIX} and \ref{fig:MICH}, which display the cumulative squared error difference between the benchmarks and the optimal indices. An upward-sloping curve indicates that the optimal selection matrix outperforms the benchmark. For the VIX, the outperformance of the optimal index remains relatively consistent during the test window, except for a notable flattening towards the end of the sample period, corresponding to a lower volatility period after the initial spike due to the COVID-19 pandemic. For inflation expectations, the outperformance of the optimal index is evident from 2016 to mid-2018, followed by a slight underperformance until 2020. Interestingly, the optimal index exhibits superior performance as inflation expectations rise towards the end of the sample period.

\insertfloat{Figure~\ref{fig:VIX} and Figure~\ref{fig:MICH}}

\section{Conclusion}
\label{sec:conclusion}

In this research, we address a critical aspect often overlooked in economic and financial analysis using textual data---the text selection process. We introduce an algorithm that determine which set of texts, among a large corpus, leads to a text-based index that is optimal for a specific objective---typically, an index that maximizes the contemporaneous relation or the predictive performance with respect to a target variable, such as inflation.  Our approach, based on a genetic algorithm and tailored crossover and mutation operators, offers a data-driven and systematic text selection procedure.

We illustrate the relevance of our approach using a large collection of news articles from the Wall Street Journal. In particular, we show how to improve the existing news-based VIX index by \citet{baker2019policy} or the news-based inflation expectations index by \citet{pfajfar2013news} and \citet{marcellino2022demand}. 

\section*{Supplementary Materials}

\noindent
Please contact the corresponding author to have access to the computer code and the data sets.

\section*{Acknowledgements}

\noindent
We are grateful to IVADO and the Natural Sciences and Engineering Research Council of Canada (grant RGPIN-2022-03767) for their financial support.

\newpage
\singlespacing

\begin{table}[H]
\centering
\caption{\textbf{Optimization Process for the EPU Index Target}\\
This table reports the number of active tokens, the training, validation, and test window root mean square error (RMSE, $\times 100$) for each optimization epoch. We report this for $K=1,2,3$-dimensional selection matrix optimization.} 
\label{tab:epu}
\scalebox{0.83}{
\begin{tabular}{lcccc}
\toprule
\multicolumn{5}{c}{Panel A: $K = 1$}  \\ 
\midrule
Epoch & \#Active  & Training & Validation & Test \\ 
\midrule
1 & 1 & 49.94 & 58.07 & 51.73 \\ 
2 & 1 & 49.94 & 58.07 & 51.73 \\ 
3 & 3 & 47.78 & 56.97 & 53.61 \\ 
4 & 3 & 47.78 & 56.97 & 53.61 \\ 
5 & 5 & 43.73 & 55.64 & 60.66 \\ 
6 & 5 & 43.73 & 55.64 & 60.66 \\ 
7 & 5 & 43.73 & 55.64 & 60.66 \\ 
8 & 5 & 43.73 & 55.64 & 60.66 \\ 
9 & 5 & 43.73 & 55.64 & 60.66 \\ 
10 & 5 & 43.73 & 55.64 & 60.66 \\ 
11 & 5 & 43.73 & 55.64 & 60.66 \\ 
12 & 5 & 43.73 & 55.64 & 60.66 \\ 
13 & 5 & 43.73 & 55.64 & 60.66 \\ 
14 & 5 & 43.73 & 55.64 & 60.66 \\ 
15 & 5 & 43.73 & 55.64 & 60.66 \\ 
Pruning & 4 & 46.67 & 55.77 & 57.61 \\ 
\midrule
\multicolumn{5}{c}{Panel B: $K = 2$}  \\ 
\midrule
Epoch & \#Active  & Training & Validation & Test \\ 
\midrule
1 & 2 & 50.92 & 56.55 & 43.74 \\ 
2 & 2 & 50.92 & 56.55 & 43.74 \\ 
3 & 4 & 44.88 & 56.20 & 45.60 \\ 
4 & 5 & 44.47 & 55.40 & 46.98 \\ 
5 & 6 & 43.49 & 55.21 & 47.64 \\ 
6 & 7 & 41.83 & 53.41 & 47.16 \\ 
7 & 8 & 41.35 & 53.23 & 47.28 \\ 
8 & 9 & 39.51 & 48.06 & 54.53 \\ 
9 & 10 & 38.67 & 47.66 & 54.44 \\ 
10 & 10 & 38.67 & 47.66 & 54.44 \\ 
11 & 12 & 33.36 & 41.73 & 53.69 \\ 
12 & 12 & 33.36 & 41.73 & 53.69 \\ 
13 & 12 & 33.36 & 41.73 & 53.69 \\ 
14 & 12 & 33.36 & 41.73 & 53.69 \\ 
Pruning & 12 & 33.36 & 41.73 & 53.69 \\ 
\midrule
\multicolumn{5}{c}{Panel C: $K = 3$}  \\ 
\midrule
Epoch & \#Active  & Training & Validation & Test \\ 
\midrule
1 & 3 & 56.57 & 62.70 & 47.57 \\ 
2 & 4 & 49.35 & 57.40 & 48.58 \\ 
3 & 5 & 47.65 & 53.26 & 47.71 \\ 
4 & 6 & 47.13 & 52.45 & 47.02 \\ 
5 & 7 & 42.78 & 46.36 & 45.10 \\ 
6 & 8 & 41.11 & 43.90 & 44.90 \\ 
7 & 9 & 33.72 & 37.51 & 40.81 \\ 
8 & 10 & 11.37 & 12.26 & 17.34 \\ 
9 & 11 & 0.00 & 0.00 & 0.00 \\ 
10 & 11 & 0.00 & 0.00 & 0.00 \\ 
11 & 11 & 0.00 & 0.00 & 0.00 \\ 
12 & 11 & 0.00 & 0.00 & 0.00 \\ 
13 & 11 & 0.00 & 0.00 & 0.00 \\ 
Pruning & 10 & 0.00 & 0.00 & 0.00 \\ 
\hline
\end{tabular}}
\end{table}

\begin{table}[H]
\caption{\textbf{Performance Results of Selection Matrix Process for the VIX and Inflation Targets}\\
This table reports the root mean squared error (RMSE, $\times 100$) measure (under various time windows) for the optimal selection matrix as 
well as for its benchmark for the VIX and MICH (twelve-month inflation expectations).}
\label{tab:vix}
\centering
\begin{tabular}{lccc}
\toprule
\multicolumn{4}{c}{Panel A: VIX}\\
\midrule
& Training & Validation & Test \\ 
\midrule
Optimal & 31.53 & 43.93 & 46.14 \\ 
EMV & 61.10 & 71.73 & 72.78 \\ 
\midrule
\multicolumn{4}{c}{Panel B: MICH}\\
\midrule
& Training & Validation & Test \\ 
\midrule
Optimal & 52.64 & 64.28 & 66.16 \\ 
Inflation & 68.09 & 87.89 & 87.73 \\ 
\bottomrule
\end{tabular}
\end{table}

\newpage

\newpage

\end{document}